\documentclass[aps,preprint]{revtex4}%
\usepackage{amsfonts}
\usepackage{amsmath}
\usepackage{amssymb}
\usepackage{graphicx}%
\begin{document}
\title{{\LARGE A GENERAL RELATIVISTIC ROTATING EVOLUTIONARY UNIVERSE - PART II}}
\author{Marcelo Samuel Berman$^{1}$}
\affiliation{$^{1}$Instituto Albert Einstein/Latinamerica\ - Av. Candido Hartmann, 575 -
\ \# 17}
\affiliation{80730-440 - Curitiba - PR - Brazil - email: msberman@institutoalberteinstein.org}
\date{(Original: 26 January, 2008. Last Version: 18 March, 2008).}

\begin{abstract}
As a sequel to (Berman, 2008a), we show that the rotation of the Universe can
be dealt by generalised Gaussian metrics, defined in this paper.
Robertson-Walker's metric has been \ employed with proper-time, in its
standard applications; the generalised Gaussian metric imply in the use of a
non-constant temporal metric coefficient modifying Robertson-Walker's standard
form. Experimental predictions are made.

\end{abstract}
\maketitle

\begin{center}

{\LARGE A GENERAL RELATIVISTIC ROTATING EVOLUTIONARY UNIVERSE - PART II}

\bigskip Marcelo Samuel Berman
\end{center}

\bigskip{\LARGE 1. Introduction}

Standard aspects concerning the treatment of rotation in General Relativity
(GRT), were outlined by Islam (1985). Rotation of the Universe was dealt by
Berman from several angles, like the Machian (Berman, 2007; 2007a; 2007b;
2007c) and under the Robertson-Walker's metric (Berman, 2008; 2008a). We now
shall be dealing with the cosmological metric that would be fitted for the
study of a rotating and expanding Universe, in a parallel approach with new
kinds of arguments. 

\bigskip

It is usual and practical, to employ metrics "adapted" to solve a given
physical problem. Any metric induces a topology. The same topology,
nevertheless, can be met by other less adapted metrics. The whole set of
metrics, can be used to solve any particular problem. According to Synge
(1960), what matters is only curvature in each given point, and not, for
instance, a typical observer's proper quadri-acceleration along his
world-line, for some metric. Joshi (1993) comments that GRT restricts
spacetime, according to the principle of local flatness and its Special
Relativistic framework. However, local restrictive physics, do not affect
global topology of the spacetime manifold. Boundary conditions, to the
contrary, limit the possible topologies. One such condition is revealed by
Mach's Principle, and may satisfy macroscopical approaches, but for
microphysics, a more elaborated topology must be found. In a private
communication, A. Sant'Anna, has pointed out that micro-topologies are now
under research, in order to produce viable scenarios in Quantum computer theory.

\bigskip\bigskip

Without delving in the above related details, I showed elsewhere, that for a
GRT rotating model, Robertson-Walker's metric could be generalised in order to
fit the problem of rotation plus expansion. This will be the subject of our
present paper.

\bigskip

{\LARGE 2. Rotating Evolutionary Metrics}

\bigskip

When we are not working with proper time \ $\tau$\ \ , but make a
transformation to any other time-coordinate \ $t$\ \ . Generally, instead, we write:

\bigskip

$d\tau^{2}\equiv g_{00}(r,\theta,\phi,t)$ $dt^{2}$ \ \ \ \ \ \ \ \ \ \ \ \ \ . \ \ \ \ \ \ \ \ \ \ \ \ \ \ \ \ \ \ \ \ \ \ \ \ \ \ \ \ \ \ \ \ \ \ \ \ \ \ \ \ \ \ \ \ \ (1)

\bigskip

The Robertson-Walker's metric is written usually in terms of proper time, namely,

\bigskip

$ds^{2}=d\tau^{2}-\frac{R^{2}}{\left(  1+k\frac{r^{2}}{4}\right)  ^{2}}%
d\sigma^{2}$ \ \ \ \ \ \ \ \ \ \ \ \ \ . \ \ \ \ \ \ \ \ \ \ \ \ \ \ \ \ \ \ \ \ \ \ \ \ \ \ \ \ \ \ \ \ \ \ \ \ \ \ \ \ \ \ (2)

\bigskip

If we change from the angular coordinate \ $\phi$\ \ \ towards \ \ $\tilde
{\phi}$\ \ , such that:

\bigskip

$\tilde{\phi}=\phi-\omega$ $t$\ \ \ \ \ \ \ \ \ \ \ \ \ \ \ \ \ \ \ , \ \ \ \ \ \ \ \ \ \ \ \ \ \ \ \ \ \ \ \ \ \ \ \ \ \ \ \ \ \ \ \ \ \ \ \ \ \ \ \ \ \ \ \ \ \ \ \ \ \ \ \ \ \ \ (3)

\bigskip

or,

\bigskip

$d\tilde{\phi}=d\phi-\omega$ $dt$\ \ \ \ \ \ \ \ \ \ \ \ \ \ \ \ \ \ \ , \ \ \ \ \ \ \ \ \ \ \ \ \ \ \ \ \ \ \ \ \ \ \ \ \ \ \ \ \ \ \ \ \ \ \ \ \ \ \ \ \ \ \ \ \ \ \ \ \ \ (3a)

\bigskip

we shall \ find that tri-dimensional metric element becomes,

\bigskip

$d$ $\tilde{\sigma}^{2}=dr^{2}+r^{2}d\theta^{2}+r^{2}\sin^{2}\theta\left(
d\phi-\omega dt\right)  ^{2}=$

\bigskip\ \ \ \ \ \ \ \ \ \ \ \ \ \ \ \ \ \ \ \ \ \ \ \ \ \ \ \ \ \ \ \ \ \ \ \ \ \ \ \ \ \ \ \ \ \ \ \ \ \ \ \ \ \ \ \ \ \ \ \ \ \ \ \ \ \ \ \ \ \ \ \ \ \ \ \ \ \ \ \ \ \ \ \ \ \ \ \ \ \ \ \ \ \ \ \ (4)

$=d\sigma^{2}+r^{2}\sin^{2}\theta$ $\omega^{2}dt^{2}-2\omega$ $r^{2}\sin
^{2}\theta$ $d\phi$ $dt$\ \ \ \ \ \ \ \ \ \ \ \ \ \ \ .\ \ \ \ \ 

\bigskip

\bigskip The new metric is of the following form:

\bigskip

$d$ $\tilde{s}^{2}=g_{00}dt^{2}-\frac{R^{2}}{\left(  1+k\frac{r^{2}}%
{4}\right)  ^{2}}\left[  d\sigma^{2}-2\omega r^{2}\sin^{2}\theta d\phi
dt\ \right]  $\ \ \ \ \ \ \ , \ \ \ \ \ \ \ \ \ \ \ \ \ \ \ \ \ \ \ (5)

\bigskip

with,

\bigskip

$g_{00}=1-\frac{\omega^{2}R^{2}r^{2}\sin^{2}\theta}{\left(  1+k\frac{r^{2}}%
{4}\right)  ^{2}}$ \ \ \ \ \ \ \ \ \ \ \ \ \ \ \ \ . \ \ \ \ \ \ \ \ \ \ \ \ \ \ \ \ \ \ \ \ \ \ \ \ \ \ \ \ \ \ \ \ \ \ \ \ \ \ \ \ \ \ \ \ \ \ (5a)

\bigskip

Consider now a totally comoving observer: his proper time is given by,

\bigskip

$d\tau^{2}=g_{00}dt^{2}=\left[  1-\varpi^{2}R^{2}\right]  $ $dt^{2}$
\ \ \ \ \ \ \ \ \ \ \ \ \ , \ \ \ \ \ \ \ \ \ \ \ \ \ \ \ \ \ \ \ \ \ \ \ \ \ \ \ \ \ \ \ \ \ \ (6)

\bigskip

where \ 

$\bigskip$

$\varpi=\frac{\omega r\sin\theta}{\left(  1+k\frac{r^{2}}{4}\right)  }%
$\ \ \ \ \ \ \ \ \ \ \ \ \ \ \ \ \ \ . \ \ \ \ \ \ \ \ \ \ \ \ \ \ \ \ \ \ \ \ \ \ \ \ \ \ \ \ \ \ \ \ \ \ \ \ \ \ \ \ \ \ \ \ \ \ \ \ \ \ \ \ \ \ \ \ \ (7)

\bigskip

If we \ want to preserve \ $g_{00}>0$\ \ \ , so that equation (6) represents
real proper time, we may solve the problem with,

\bigskip

$\varpi=\frac{\alpha}{R}$\ \ \ \ \ \ \ \ \ , \ \ \ \ \ \ \ \ \ \ \ \ \ \ \ \ \ \ \ \ \ \ \ \ \ \ \ \ \ \ \ \ \ \ \ \ \ \ \ \ \ \ \ \ \ \ \ \ \ \ \ \ \ \ \ \ \ \ \ \ \ \ \ \ \ \ \ \ \ \ \ \ \ (8)

\bigskip

with \ $\alpha^{2}<1$\ \ \ \ .

\bigskip

For instance, we could fix \ $\dot{\alpha}=0$\ \ so that, \ $\alpha
=\alpha\left(  r\right)  $\ \ then we would also find,

\bigskip

$\omega=\frac{\beta}{R}$\ \ \ \ \ \ \ \ \ \ \ \ \ \ \ \ \ ,

\bigskip

with a similar condition, \ $\dot{\beta}=0$\ \ , and \ \ $\beta=\beta(r)$\ \ \ .\ 

\bigskip

We call \ $\varpi$\ \ as the effective angular speed of the Universe; it has a
striking similarity with Berman's solution for a Machian rotating Universe
(Berman, 2007b). We notice that such angular speed causes a kind of
centripetal ubiquitous acceleration having a universal character and which
caused the so-called Pioneer anomaly, in two space probes, launched by NASA
more than thirty years ago. By the same token, we define a tangential
\ \ speed ,\ $\ \ $

$\bigskip$

$\ \ V=\varpi$ $R<1$\ \ .

\bigskip

In order to make contact with usual cosmological theory, we consider now that,
the non-diagonal metric term which points towards a Universal precession of
gyroscopes, is in the same token, as in the Lense-Thirring metric. Consider
now a semi-comoving observer, defined by the condition \ \ $d$ $\tilde{\phi
}=0$\ \ . If we write,

\bigskip

$\omega=\frac{d\phi}{dt}$ \ \ \ \ \ \ \ \ \ \ \ \ \ ,

\bigskip

the non-diagonal term becomes diagonalized, \ like,

\bigskip

$2\frac{\varpi^{2}}{\omega}R^{2}d\phi$ $dt=2\varpi^{2}R^{2}dt^{2}%
$\ \ \ \ \ \ \ \ \ \ \ \ \ \ \ \ \ . \ \ \ \ \ \ \ \ \ \ \ \ \ \ \ \ \ \ \ \ \ \ \ \ \ \ \ \ \ \ \ \ \ \ \ \ \ \ \ \ \ \ \ \ \ (9)

\bigskip

In this case, the apparent temporal metric coefficient becomes,

\bigskip

$(g_{00})_{ap}=1-\varpi^{2}R^{2}+2\varpi^{2}R^{2}=1+\varpi^{2}R^{2}$
\ \ \ \ \ \ \ \ \ \ . \ \ \ \ \ \ \ \ \ \ \ \ \ \ \ \ \ \ \ \ \ \ \ \ \ \ (10)\ 

\bigskip

With the value of \ $\varpi$\ \ given by relation (8) the apparent temporal
metric coefficient becomes time-independent. Of course, the real temporal
coefficient is then also so.

\bigskip

{\LARGE 3. Generalised Gaussian Metrics}

\bigskip

We refer to Berman (2007; 2007a), for a presentation of the Gaussian metrics
which does not suffer from the "trap" of considering \ \ $g_{00}=$\ \ constant
\ \ \ for their definition. What really matters in a Gaussian metric, is that
the time axis is orthogonal to the tri-space. When such metric represents a
tri-space which is not into rotation around the time axis, \ coordinate time
means proper time \ \ $\tau$\ \ \ . Now relax the restriction, and consider
that the orthogonal tri-space is rotating around the time axis: the time
coordinate that we must use is \ $t$\ \ \ , defined by \ (1) . This was indeed
shown in the previous section, and points to what we shall call Generalised
Gaussian metrics.

\bigskip

Berman's definition of Gaussian metrics (Berman 2007; 2007a), constrains only
the metric by the condition,

\bigskip

\ $g^{ij}\frac{\partial g_{i0}}{\partial t}=0$%
\ \ \ \ \ \ \ \ \ \ \ \ \ \ ($i,j=1,2,3$) \ . \ \ \ \ \ \ \ \ \ \ \ \ \ \ \ \ \ \ \ \ \ \ \ \ \ \ \ \ \ \ \ \ \ \ \ \ \ \ \ (11)

\bigskip

On the other hand, \ Gaussian normal coordinates, are defined by the condition,

\bigskip

$g_{i0}(t)=0$ \ \ \ \ \ \ \ \ \ \ \ \ \ \ . \ \ \ \ \ \ \ \ \ \ \ \ \ \ \ \ \ \ \ \ \ \ \ \ \ \ \ \ \ \ \ \ \ \ \ \ \ \ \ \ \ \ \ \ \ \ \ \ \ \ \ \ \ \ \ \ \ \ \ \ \ \ \ (12)

\bigskip

Then, the following condition applies for a comoving observer:

\bigskip

$g_{00}u^{0}=1$ \ \ \ \ \ \ \ \ \ \ \ \ \ \ \ , \ \ \ \ \ \ \ \ \ \ \ \ \ \ \ \ \ \ \ \ \ \ \ \ \ \ \ \ \ \ \ \ \ \ \ \ \ \ \ \ \ \ \ \ \ \ \ \ \ \ \ \ \ \ \ \ \ \ \ \ \ \ (13)

\bigskip

where, \ \ \ $u^{0}$\ \ \ is the temporal component of the quadri-velocity.

\bigskip

From the cosmological point of view, it has been suggested that rotation of
the Universe is associated with cosmic microwave background radiation's
quadrupole anisotropies. These have not been significant: this may be
attributed to low angular speeds, less than what is possibly measured by
present technology. We must remember that the cosmic no-hair conjecture is
really established by means of an inflationary phase erasing the angular speed
(remember equation (8), with \ $R$\ \ exponentially increasing).

\bigskip

Another point is that CMBR analyses, only apply to the equation of null
geodesics, \ $ds=0$\ \ .\ \ To the contrary, the Pioneer anomaly deals with
\ $ds\neq0$\ \ . It must be stressed that a variable temporal metric
coefficient has been studied long ago by Gomide and Uehara (1981).

\bigskip

{\LARGE 4. Final Comments}

\bigskip

The rotation of the Universe, not only explains the Pioneers' anomaly, but
would be in the right direction, in order to explain the left handed
preference of neutrinos' spins, parity violations and the related
matter-antimatter asymmetry (Feynman et al., 1965). \bigskip It has been found
that the DNA helix is left handed. Our bodies are not symmetric; molluscs have
likewise shells; aminoacids in living bodies, too (Barrow and Silk, 1983).

\bigskip

We predict that chaotic phenomena, and fractals, in the Universe, as well as
rotations of Galaxies and their clusters, must have a predilection towards the
left hand. We predict that the directions of the magnetic field of the
Universe, and rotation, are related with the laws of electrodynamics and the left-hand.

\bigskip It is important to note the preliminary results on the Universe
rotation, by Birch (1982; 1983) and Gomide, Berman and Garcia (1986). Sciama's
inertial theory (Sciama, 1953), also contemplated a rotation speed of the type
given by us (see (8)). The spin of the Universe is a subject of two recent,
and a forthcoming, papers by Berman (2007b; 2007c; 2008). Machian rotations
are dealt by Berman (2007; 2007a).

\bigskip

\bigskip I predict that with improving technological tools, the rotation of
the Universe will be experimentally measured in the future.

\bigskip

{\LARGE Acknowledgements}

\bigskip

The author thanks his intellectual mentors, Fernando de Mello Gomide and M. M.
Som, and also to Marcelo Fermann Guimar\~{a}es, Nelson Suga, Mauro Tonasse,
Antonio F. da F. Teixeira, and for the encouragement by Albert, Paula and Geni.

\bigskip

\bigskip{\Large References}

\bigskip Barrow, J.D; Silk, J. (1983) - \textit{The Left Hand of Creation: The
Origin and Evolution of Expanding Universe, }Basic Books, New York.

Berman,M.S. (2007) - \textit{Introduction to General Relativistic and Scalar
Tensor Cosmologies}, Nova Science, New York.

Berman,M.S. (2007a) - \textit{Introduction to General Relativity and the
Cosmological Constant Problem}, Nova Science, New York.

Berman,M.S. (2007b) - \textit{The Pioneer Anomaly and a Machian Universe,
}Astrophysics and Space Science, \textbf{312}, 275.

\bigskip Berman,M.S. (2007c) - \textit{Shear and Vorticity in Inflationary
Brans-Dicke Cosmology with Lambda-Term, }Astrophysics and Space Science,
\textbf{310, }205.

Berman,M.S. (2008) - \textit{Shear and Vorticity in a Combined
Einstein-Cartan-Brans-Dicke Inflationary Lambda-Universe, }Astrophysics and
Space Science, to appear. For an earlier version, see Los Alamos Archives:
http://arxiv.org/abs/physics/0607005 

Berman,M.S. (2008a) - \textit{A General Relativistic Rotating Evolutionary
Universe, }Astrophysics and Space Science, to appear. For an earlier version,
see Los Alamos Archives: http://arxiv.org/abs/physics/0712.0821 [physics.gen-ph]

Birch, P. (1982) - Nature, \textbf{298}, 451.

Birch, P. (1983) - Nature, \textbf{301}, 736.

\bigskip Feynman, R.P.; Leighton, R.B.; Sands, M. (1965) - \textit{The Feynman
Lectures on Physics.} Volume 1, Addison-Wesley, Reading.

Gomide,\ F.M.; Berman, M.S.; Garcia, R.L. (1986) - Rev. Mex. AA; \textbf{12}, 46.

Gomide,\ F.M.; Uehara, M. (1981) - Astronomy and Astrophysics, \textbf{95}, 362.

Islam, J.N. (1985) - \textit{Rotating Fields in General Relativity,} Cup, Cambridge.

Joshi, P.S. (1993) - \textit{Global Aspects in Gravitation and Cosmology},
Oxford University Press, Oxford.

Sciama, D.N. (1953) - MNRAS, \textbf{113}, 34.

Synge, J.L. (1960) - \textit{Relativity: The General Theory,} North-Holland, Amsterdam.

\end{document}